\documentclass[preprint]{elsarticle}

\usepackage{lineno,hyperref} 
\usepackage{bm}
\usepackage{verbatim}
\usepackage{color}
\usepackage{lineno}
\usepackage{amssymb}
\usepackage{hyperref}
\usepackage{eufrak}
\usepackage{tabularx}
\usepackage{amsmath}
\usepackage{pifont}
\newcommand{\cmark}{\text{\ding{51}}}
\newcommand{\xmark}{\text{\ding{55}}}
\modulolinenumbers[5]

\begin{document}

\title{Electromagnetic wave propagation in spatially homogeneous yet smoothly time-varying dielectric media}

    \author[mpipks]{Armen G.~Hayrapetyan\corref{cor1}}
    \author[mpipks]{J\"org B.~G\"otte}
    \author[ysu]{Karen K.~Grigoryan}
    \author[hi,fsuj]{Stephan Fritzsche}
    \author[ysu]{Rubik G.~Petrosyan}
    
    \cortext[cor1]{Corresponding author \\
    \textit{Email address:} armen@pks.mpg.de (Armen G.~Hayrapetyan)}
    
    \address[mpipks]{Max-Planck-Institut f\"ur Physik komplexer Systeme, N\"othnitzer Str. 38, 01187 \\ Dresden, Germany}
    \address[ysu]{Yerevan State University, 1 Alex Manoogian Str., 0025 Yerevan, Armenia}
    \address[hi]{Helmholtz-Institut Jena, 07743 Jena, Germany} 
    \address[fsuj]{Theoretisch-Physikalisches Institut, Friedrich-Schiller-Universit\"at Jena, 07743 \\ Jena, Germany}

\begin{abstract}

We explore the propagation and transformation of electromagnetic waves through spatially homogeneous yet 
\textit{smoothly time}-dependent
media within the framework of classical electrodynamics. By modelling the smooth transition, 
occurring during a finite period $\tau$,
as a phenomenologically realistic and sigmoidal change of the dielectric permittivity,
an analytically exact solution to Maxwell's equations is derived for the 
electric displacement in 
terms of hypergeometric functions. Using this solution, we show the possibility of 
\textit{amplification} and \textit{attenuation} of waves and associate this with the 
\textit{decrease} and \textit{increase} of the time-dependent permittivity.
We demonstrate, moreover, that such an energy exchange between waves and non-stationary media 
leads to the transformation (or conversion) of frequencies. Our results may pave the way towards 
controllable light-matter interaction in time-varying structures.

\end{abstract}


\maketitle

\section{Introduction}

Controlling the optical properties of photonic structures has been a topic of significant interest
for last few decades both in fundamental~\cite{Boyd:08,Joannopoulos:08} and applied 
research~\cite{Hulin:86,Jewell:89,Rivera:94}. Recent advances 
in technology and instrumentation have made it possible to realize such control systems 
via ultrafast switching of the \textit{time-dependent} dielectric permittivity (or the refractive 
index)~\cite{Yamamoto:98,Nakamura:01,Neilsen:06,Harding:07,Euser:08,Nozaki:10,Ctistis:11}. 
The two most relevant mechanisms of modifying the permittivity in the time domain
are the excitation of the so-called free charge-carriers and the electronic
Kerr effect. While the free-carriers
are typically induced by strong pump pulses creating electron-hole 
pairs in semiconductors~\cite{Leonard:02,Mazurenko:03,Bristow:03,Euser:05}, 
the Kerr effect arises from the non-linear (instantaneous) response of bound electrons
to the applied field~\cite{Boyd:08,Inouye:03,Hu:03,Yuece:13}. 
These mechanisms enable one to employ
the optical switching of the refractive index for various purposes, such as 
quantum interference~\cite{Harris:98}, information 
processing~\cite{Dawes:05,Bajcsy:09,Venkataraman:10}, 
material science~\cite{Yoshimura:02,Srinivas:08,Tajima:12},
control of spontaneous emission~\cite{Thyrrestrup:13}, 
and several others~\cite{Johnson:02,Fushman:07,Sivan:11,Harding:12,Xia:13,Sivan:15}.

The investigation of dynamics of 
both optical and matter \textit{waves} in instantaneously time-varying 
structures have also been in the focus of intense research 
throughout the last decades.
Until now, several effects have already been proposed, for instance, 
to account for the stimulated electron-light interaction~\cite{Avetisyan:78} and 
photon generation~\cite{Cirone:97}, 
to investigate irregular alternation of phases of electromagnetic waves~\cite{Nerukh:99} and
the energy exchange between waves and non-stationary media~\cite{Mkrtchyan:10}.
In other scenarios, modulation and conversion of frequencies as well as the 
transformation of waves have been explored in non-stationary waveguides, resonators and plasmas~[cf.~books~\cite{Ginzburg:79,Ginzburg:90,Nerukh:12}
and references therein].
The idea of looking at such phenomena
stems from the pioneering papers by Morgenthaler~\cite{Morgenthaler:58}
and Ginzburg and Tsytovich~\cite{Ginzburg:73,Ginzburg:74}, who
considered the velocity modulation of waves and
the transition radiation of charged particles in time-dependent environments. 
Recently, the dynamics of sound waves have also been examined in non-stationary fluids
with either a sudden~\cite{Mkrtchyan:10} or smooth~\cite{Hayrapetyan:11,Hayrapetyan:13} change of 
the medium, leading to the frequency conversion of waves
that have already become accessible in an experiment~\cite{Wang:15}.

In most of the studies related to the transformation of electromagnetic waves, 
the parameters describing the medium are assumed to vary sharply (or step-like) with time, i.e., when
the time duration during which the medium experiences a change is much shorter than the
propagation period of the wave, $\tau \ll 2 \pi / \omega$.
Although such an assumption simplifies the theory and describes the relevant effects,
it rarely describes the reality of switching processes and should be replaced with a smooth transition. This is especially important for experiments in which the switching duration is comparable to the period of light, so that $\tau \lesssim  2 \pi / \omega$.

In this paper, we re-visit the problem of transformation of electromagnetic 
waves in time-dependent media and provide a step forward towards understanding of the impact 
of non-stationary environments, \textit{continuously} changing in time,
on the dynamical properties of waves. 
To this end, we examine the propagation and transformations of waves in 
time-varying and, at the same time, spatially homogeneous (i.e., uniform) dielectric media.
A particular emphasis is placed on studying how a \textit{smoothly} time-dependent 
dielectric permittivity affects the energy (flux) and the frequency of transformed waves. 
In view of this, 
we derive a generalized wave equation for the electric displacement and 
obtain an analytically exact solution expressed via hypergeometric functions
for a judiciously chosen  
sigmoidal change of the permittivity, explicitly accounting for the finite transition period $\tau$.
Using this solution, we show that an energy exchange occurs between electromagnetic waves
and non-stationary media, which is further
demonstrated to lead to either amplification or attenuation of waves depending on whether the 
refractive index decreases or increases as a function of time. For a monochromatic
incident wave, moreover, such an energy \textit{non}-conservation gives rise to the
transformation (or conversion) of frequency due to the (more or less) abrupt change of the refractive index,
quite similar to that of the sound wave frequency~\cite{Hayrapetyan:11,Hayrapetyan:13}.

The paper is organized as follows. In the next section, we derive a generalized 
wave equation for the electric displacement when the time-varying dielectric permittivity 
remains uniform in space. 
While continuity conditions for the electric displacement 
and magnetic induction are used to account for the sudden transition (Subsection~\ref{sudden.change}), 
rigorous exact solutions to the time-dependent wave equation 
are obtained in the case of the smooth transition (Subsection~\ref{smooth.change}).
These solutions are then discussed in Section \ref{res.disc}
and a few effects are predicted, such as the energy exchange, 
amplification and attenuation as well as frequency conversion of waves.
In Section~\ref{conclusion}, we conclude with future research directions.

\section{Theory of time-dependent propagation and transformation of electromagnetic waves}
\label{theory}

In order to describe the temporal dynamics of light in 
spatially homogeneous and isotropic media, we 
start from the source-free Maxwell equations in Gaussian units
\begin{eqnarray}
\label{Maxwell1a}
	\bm\nabla \times \bm{H} & = & \frac{1}{c} \frac{\partial \bm{D}}{\partial t} \, , \\
\label{Maxwell1b}
	\bm\nabla \times \bm{E} & = & - \frac{1}{c} \frac{\partial \bm{H}}{\partial t} \, , \\
\label{Maxwell1c}	
	\bm \nabla \cdot \bm{D} & = & 0 \, , \\
\label {Maxwell1d}
		\bm \nabla \cdot \bm{H} & = & 0 \, ,
\end{eqnarray}
where $\bm \nabla$ is the vector differential operator, 
``cross'' and ``dot'' mean vector and inner products, respectively. 
Maxwell's equations self-consistently characterize the electromagnetic field state
only in a vacuum. In a general medium, however, constitutive relations must be added to 
Eqs.~(\ref{Maxwell1a}-\ref{Maxwell1d}) to provide a complete description of waves~\cite{Landau:84}.
For an isotropic medium, 
the electric field $\bm E$ and displacement $\bm D$  
are related via the standard constitutive relation $\bm D = \varepsilon \bm E$, with $\varepsilon$ 
being the scalar dielectric permittivity.
For non-magnetic media, moreover, the magnetic permeability $\mu = 1$
as in most of the experiments on optical switching, 
so that the magnetic field $\bm H$ equals to the induction $\bm B$.
In this particular case, therefore, 
the dielectric permittivity $\varepsilon$ and the
refractive index $n$ are related by a simple formula $\varepsilon = n^2$.

For time-varying and \textit{space}-independent media, i.e.,
when the dielectric permittivity is only a function of time, $\varepsilon \left( t \right)$,
an exact wave equation can be derived for the electric displacement 
from Eqs.~(\ref{Maxwell1a})-(\ref{Maxwell1c}) 
\begin{eqnarray}
\label{Maxwell2}
	\bm \nabla^2 \bm D \left( \bm r , t \right) - \frac{ \varepsilon \left( t \right)}{c^2} \,
	\frac{\partial^2  \bm D \left( \bm r , t \right) }{\partial t^2} & = & 0 \, .
\end{eqnarray}
Given that the medium is spatially uniform, we seek for the solution of this equation in the form 
\begin{eqnarray}
\label{Ansatz1}
	\bm D \left( \bm r , t \right)	
	\,\, \equiv \,\, \hat{\bm d} \, D \left( \bm r , t \right)
	\,\, = \,\,  \hat{\bm d} \, e^{i \bm k \cdot \bm r} \, {\cal D} \left( t \right) \, ,
\end{eqnarray}
where $\hat{\bm d}$ is a unit vector along the displacement $\bm D$, while
$\bm k$ and $\bm r$ are the wave and position vectors, respectively. 
The ansatz (\ref{Ansatz1}), which is necessary to assure time
evolution of the system, characterizes the transformation of waves and accounts for 
a modification of frequencies of transformed waves. This is reminiscent of the counterpart scenario, 
when the space dependence of the medium implies 
$\bm D \left( \bm r , t \right) = \hat{\bm d} \, e^{- i \omega t} {\cal D} \left( \bm r \right)$
for a monochromatic wave traveling with frequency $\omega $~\cite{Landau:84}.

Next, by combining Eqs.~(\ref{Maxwell2}) and (\ref{Ansatz1}), we obtain 
a one-dimensional equation for ${\cal D} \left( t \right)$
\begin{eqnarray}
\label{Maxwell3}
	\frac{d^2 {\cal D} \left( t \right)}{d t^2} + \frac{c^2 k^2}{\varepsilon \left( t \right)} \, {\cal D} \left( t \right) & = & 0 \, ,
\end{eqnarray}
where the permittivity appears in the denominator of the prefactor of ${\cal D}$,
in contrast to the position dependent wave equation
with the permittivity in the nominator~[cf., e.g., Eqs.~(88.3-4) of Ref.~\cite{Landau:84}].
Built upon the explicit form of
$\varepsilon \left( t \right)$, Eq.~(\ref{Maxwell3}) allows solutions
which characterize dynamics of waves in non-stationary media
\textit{independent} of the nature of switching or tuning of the permittivity.
Moreover, an analogous generalized equation for sound waves can also be
derived from Euler's and continuity equations in non-stationary fluids
[cf.~Eq.~(34) of Ref.~\cite{Hayrapetyan:13}].

In the following, we shall investigate solutions of Eq.~(\ref{Maxwell3}) 
for two distinct cases: when the time-dependent dielectric permittivity experiences either sudden or smooth
change. On each of these scenarios, we shall (i) apply continuity conditions for electric displacement
and magnetic induction or (ii) rigorously solve the differential equation
with a phenomenologically realistic sigmoidal change of the permittivity,
judiciously chosen to qualitatively coincide with experimentally determined behaviour.
Our approach is based on the direct and exact integration of the time-dependent wave equation,
in contrast to other theoretical methods, such as Green's functions representation~\cite{Felsen:70}, 
the Wentzel-Kramers-Brillouin-Jeffreys approximation~\cite{Fante:71,Averkov:80} or the 
Volterra integral equation approach~\cite{Fedotov:03,Shabanov:07} 
(see also Refs.~\cite{Nerukh:12,Mishchenko:14}).

\subsection{Sudden change of the dielectric permittivity}
\label{sudden.change}

\begin{figure}
\centering
\includegraphics[width=0.48\textwidth]{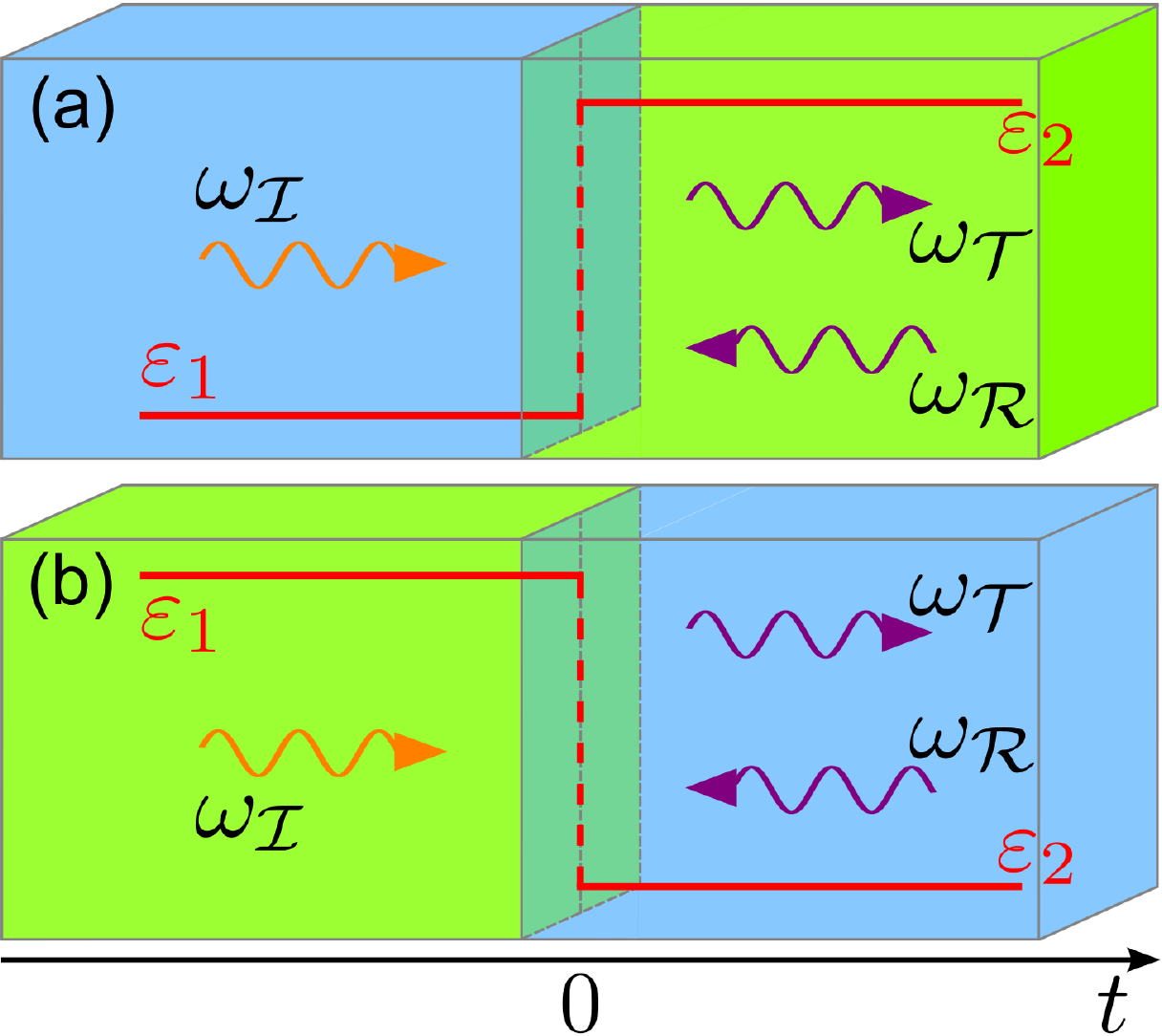}
\caption{Transformation of electromagnetic waves in spatially homogeneous dielectric media when
the permittivity experiences a sudden increase (a) or decrease (b) at time $t = t_s = 0$.
Frequencies $\omega_{\cal I}$ and $\omega_{\cal T}$ 
illustrate the propagation forward-in-time, the frequency $\omega_{\cal R}$ depicts the propagation
backward-in-time. See the text for further details.} 
\label{figure1}
\end{figure}

Before we examine the dynamics of waves in spatially uniform media 
with a smoothly changing time-dependent permittivity, 
we re-visit the case of the sudden change, as reported in Ref.~\cite{Mkrtchyan:10}.
In this scenario, it is assumed that the permittivity undergoes a discrete 
change at some time $t_s$ from $\varepsilon_1$ (for $t < t_s$) to $\varepsilon_2$
(for $t > t_s$) [cf.~Fig~\ref{figure1}]. 
For these constant values of the permittivity, the solution of the wave equation~(\ref{Maxwell3})
can be expressed in terms of plane monochromatic waves. 
For a wave with the initial frequency $\omega_{\cal I} = c k / \sqrt{\varepsilon_1}$ 
and the constant amplitude~$D_{{\cal I} 0}$
\begin{eqnarray}
\label{initial.wave}
	D_{\cal I} \left( \bm r , t \right) \,\, \equiv \,\, D_{t < t_s} \left( \bm r , t \right)
	\,\, = \,\, D_{{\cal I} 0} \,
	e^{i \left( \bm k \cdot \bm r - \omega_{\cal I} t \right)} \, ,
\end{eqnarray}
the sudden change of the permittivity gives rise to the superposition of two --
reflected (``${\cal R}$'') and transmitted (``${\cal T}$'') -- waves 
\begin{eqnarray}
\label{refl.trans.waves}
	 D_{t > t_s} \left( \bm r , t \right) \,\, = \,\,
	 D_{{\cal R} 0} \,
	e^{i \left( \bm k \cdot \bm r - \omega_{\cal R} t \right)} +
	 D_{{\cal T} 0} \,
	e^{i \left( \bm k \cdot \bm r - \omega_{\cal T} t \right)} 
\end{eqnarray}
propagating with opposite frequencies  $\omega_{\cal R} =  - c k / \sqrt{\varepsilon_2}$ and 
$\omega_{\cal T} =  c k / \sqrt{\varepsilon_2} = - \omega_{\cal R} $
and distinct amplitudes $D_{{\cal R} 0}$ and $D_{{\cal T} 0}$.
In accordance with our adopted assumption about the spatial homogeneity,
the frequencies of the initial and transmitted waves are related via the conversion relation 
\begin{eqnarray}
\label{conversion}
	\omega_{\cal I} n_1 & = & \omega_{\cal T} n_2 
\end{eqnarray}
where the constant numbers $n_1 = \sqrt{\varepsilon_1}$ and 
$n_2 = \sqrt{\varepsilon_2}$ represent the refractive indices of the medium before ($t<t_s$) and 
after ($t>t_s$) the change
[cf.~Ref.~\cite{Hayrapetyan:13} for a detailed discussion].

The reflected wave in Eq.~(\ref{refl.trans.waves}) can
be interpreted as a propagation \textit{backward-in-time}
with positive frequency $\omega_{\cal T}$ as $ - i \omega_{\cal R} t = - i \omega_{\cal T} \left( - t \right)$,
or else, propagation \textit{forward-in-time} with negative frequency $\omega_{\cal R}$
as  $ - i \omega_{\cal R} t = - i \left( -  \omega_{\cal T} \right) t $.
We can nevertheless demonstrate that the reflected wave describes an actual reflection in space
by calculating the Poynting vector of the three waves (\ref{initial.wave})-(\ref{refl.trans.waves})
\begin{eqnarray}
\nonumber
	\bm S_{\cal I} & = & \frac{c^2 \left| D_{{\cal I} 0} \right|^2}{4 \pi \omega_{\cal I} \varepsilon_1^2} \, \bm k  \,\,
	\uparrow \uparrow \,\, \bm k \, ,
\\[0.1cm]
\nonumber
	\bm S_{\cal R} & = & - \, \frac{c^2 \left| D_{{\cal R} 0} \right|^2}{4 \pi \omega_{\cal T} \varepsilon_2^2} \, \bm k  \,\,
	\uparrow \downarrow \,\, \bm k \, ,
\\[0.1cm]
\label{Poynting-vectors}
	\bm S_{\cal T} & = & \frac{c^2 \left| D_{{\cal T} 0} \right|^2}{4 \pi \omega_{\cal T} \varepsilon_2^2} \, \bm k  \,\,
	\uparrow \uparrow \,\, \bm k \, .
\end{eqnarray}
As it can be readily seen, $\bm S_{\cal R}$ is anti-parallel to the 
corresponding Poynting vectors in the initial and transmitted waves.
Moreover, the phase velocities 
\begin{eqnarray}
\label{phase-velocity}
	v_{\cal I}^{\scriptscriptstyle (ph)} \, = \, \frac{\omega_{\cal I}}{k} \, , \,\,
	v_{\cal R}^{\scriptscriptstyle (ph)} \, = \, - \frac{\omega_{\cal T}}{k} 
	\, = \, - v_{\cal T}^{\scriptscriptstyle (ph)} 
\end{eqnarray}
also indicate that the reflected wave 
propagates opposite to the propagation direction of the initial and transmitted waves.
This theoretical construct for waves is still possible
to interpret in terms of dynamical observables by taking into account the fundamental 
relation between the homogeneity of time and the conservation of energy, known as Noether's theorem~\cite{Landau:76},
as we show below.

Being a solution to the wave equation~(\ref{Maxwell2}), Eqs.~(\ref{initial.wave})-(\ref{refl.trans.waves})
allow us to reveal the behaviour of waves in non-stationary media.  
To discuss this, we shall construct 
the reflectivity and transmittivity which are correspondingly
defined as ratios of (space) averaged energy fluxes of
reflected and transmitted waves to the averaged flux of the initial wave~\cite{Mkrtchyan:10}
\begin{eqnarray}
\label{reflection-transmission}
	{\cal R} \,\, = \,\, \frac{n_2 \left| D_{{\cal R} 0} / \varepsilon_2 \right|^2}{ n_1 \left| D_{{\cal I} 0} / \varepsilon_1
	\right|^2} \, , \quad
	{\cal T} \,\, = \,\, \frac{n_2 \left| D_{{\cal T} 0} / \varepsilon_2 \right|^2}{ n_1 \left| D_{{\cal I} 0} / \varepsilon_1 \right|^2} \, .
\end{eqnarray}
A similar definition, leading to the famous Fresnel's formulae, is provided in Ref.~\cite{Landau:84}
to describe the transformation of waves in spatially inhomogeneous and time-\textit{in}dependent media.
Moreover, Refs.~\cite{Glauber:00} and \cite{Dargys:12} are dedicated to
some modified Fresnel's formulae constituting the transformation of 
resonant light and polarized matter waves, respectively. Finally,
another type of Fresnel's formulae accounting for 
a time-dependent
spatial reflection of waves from non-stationary interfaces are derived in Refs.~\cite{Fante:71,Nerukh:04}.

\begin{figure}
\centering
\includegraphics[width=0.48\textwidth]{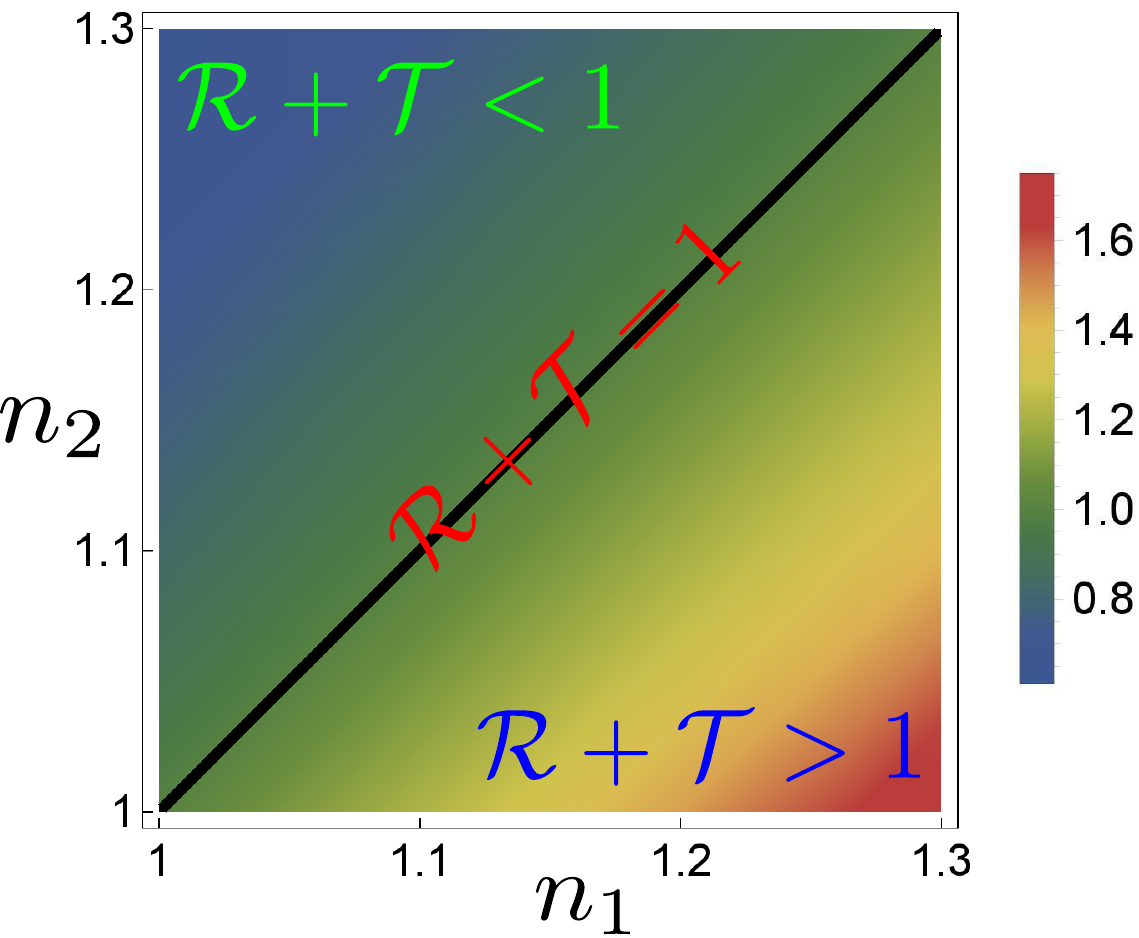}
\caption{Illustration of amplification ($n_1 > n_2$, the right triangle) and attenuation 
($n_1 < n_2$, the left triangle) of electromagnetic waves 
in non-stationary media with a sudden change of the refractive index. The density plot 
depicts the sum of the reflectivity and transmittivity~(\ref{energy-exchange.sudden})
as a function of refractive indices before ($n_1$) and after the change ($n_2$). The
black line represents the area ($n_1 = n_2$) where the energy of the wave is conserved.} 
\label{figure2}
\end{figure}

Using relations~(\ref{reflection-transmission}), we can establish the 
\textit{energy balance} between the transformed waves~(\ref{refl.trans.waves}) 
and the medium if we, without loss of generality, assume that the dielectric
permittivity (or the refractive index) suffer an abrupt change at the time $t_s = 0$.
For such an assumption, the continuity conditions for the electric displacement and magnetic induction  
\begin{eqnarray}
\nonumber
	\bm D_{t < 0} \left( \bm r , 0 \right) & = & \bm D_{t > 0} \left( \bm r , 0 \right) \, ,
\\
\label{continuity.cond}
	\bm B_{t < 0} \left( \bm r , 0 \right) & = & \bm B_{t > 0} \left( \bm r , 0 \right) \, ,
\end{eqnarray}
lead to the Fresnel-type formulae in the time domain~\cite{Mkrtchyan:10}
\begin{eqnarray}
\label{reflection.sudden}
	{\cal R} & = & \frac{n_1 \left( n_1 - n_2 \right)^2}{ 4 n_2^3} 
	\,\, = \,\, \frac{\omega_{\cal T} \left(\omega_{\cal T} - \omega_{\cal I} \right)^2}{4 \omega_{\cal I}^3 } \, , 
\\[0.2cm]
\label{transmission.sudden}
	{\cal T} & = & \frac{n_1 \left( n_1 + n_2 \right)^2}{ 4 n_2^3} 
	\,\, = \,\, \frac{\omega_{\cal T}  \left(\omega_{\cal T} + \omega_{\cal I} \right)^2}{4 \omega_{\cal I}^3} \, .
\end{eqnarray}
To get a deeper insight, we add the
expressions (\ref{reflection.sudden}) and (\ref{transmission.sudden}) and find
\begin{eqnarray}
\label{energy-exchange.sudden}
	{\cal R} + {\cal T} \,\, = \,\,
	\frac{n_1 \left( n_1^2 + n_2^2 \right)}{2 n_2^3} \,\, = \,\,
	\frac{\omega_{\cal T} \left( \omega_{\cal I}^2 + \omega_{\cal T}^2 \right)}{2 \omega_{\cal I}^3} \,\, \neq \,\, 1 \, ,
\end{eqnarray}
that shows non-conservation of energy for the wave.
As the energy is conserved for the whole ``wave + medium'' system,
the expression~(\ref{energy-exchange.sudden}) can be nevertheless interpreted as 
an energy (flux) exchange between the wave and the time-dependent medium.
Depending on whether the wave propagates to optically denser ($n_2 > n_1$) or rarer ($n_2 < n_1$) medium,
it is either attenuated (${\cal R} + {\cal T}  < 1$) or amplified
(${\cal R} + {\cal T}  > 1$), as also illustrated on Fig~\ref{figure2}.
This behaviour is quite in contrast to the propagation of sound waves in 
non-stationary fluids, when the wave is only amplified 
despite the increase or decrease of the appropriate quantities, such as
distributions of the mass density and the sound 
velocity~\cite{Mkrtchyan:10,Hayrapetyan:11,Hayrapetyan:13}.

It is important to realize that in the absence of the time-dependent change of the 
permittivity, i.e., when the temporal inhomogeneity is
``switched off'' ($\varepsilon_1 = \varepsilon_2$ or $n_1 = n_2$), the reflectivity 
${\cal R}$ vanishes, while ${\cal T} = 1$,
the energy of the wave is conserved (${\cal R} + {\cal T}  = 1$, the black line on Fig~\ref{figure2})
and no frequency conversion occurs ($\omega_{\cal T} = \omega_{\cal I}$), as one would expect.

\begin{table}
  \caption{Comparison of dynamics of electromagnetic waves in time and space domains.
  Refractive indices and the reflectivity/transmittivity (for normal incidence) in the case of the spatial
  transformation are marked with upper cases $N_1$, $N_2$ and $R$, $T$
  to distinguish from the temporal transformation.}
    \label{table1}
  \begin{tabularx}{\textwidth}{lcc}
    \hline\hline 
     & Spatial homogeneity and                                  & Spatial \textit{in}homogeneity and \\
     & temporal \textit{in}homogeneity~\cite{Mkrtchyan:10} & temporal homogeneity~\cite{Landau:84} \\
    \hline 
    Reflection           & $\displaystyle {\cal R} =  \frac{n_1 \left( n_1 - n_2 \right)^2}{4 n_2^3} $ 
                               & $\displaystyle R = \left( \frac{N_1 - N_2}{N_1+N_2} \right)^2$  \\
    Transmission      & $\displaystyle {\cal T} =   \frac{n_1 \left( n_1 + n_2 \right)^2}{4 n_2^3}$ 
                               & $\displaystyle T =  \frac{4 N_1 N_2}{\left( N_1+N_2 \right)^2}$ \\
    Energy balance  & ${\cal R} +{\cal T} \neq 1$ 
                               & $R + T = 1$  \\
    \hline\hline
  \end{tabularx}
\end{table}

For a better understanding, a comparative analysis of the transformation of waves in the time domain and
of its spatial counterpart is summarized in Table~\ref{table1}.
For the sake of simplicity, we compare the Fresnel-type formulae
(\ref{reflection.sudden}) and (\ref{transmission.sudden}) with the conventional Fresnel's formulae
for the case of the normal incidence of a wave on the interface between two different spatially
homogeneous media with refractive indices $N_1$ and $N_2$. While the 
wave does \textit{not} conserve energy throughout the propagation in the \textit{non}-stationary
medium, the energy of the wave is conserved in a stationary
medium, even if it is spatially inhomogeneous. 
These energy-related effects are a direct manifestation of either violation or 
observance of Noether's conservation theorems~\cite{Landau:76}.

After this overview, in the next subsection, we extend our studies of transformation of waves
in abruptly-varying media to the case of the adiabatic change of the dielectric permittivity. In view of this, we shall
assume that the permittivity varies smoothly during some finite transition period $\tau > 0$, which 
also plays the role of the switching duration. 
Such a smoothness is then modelled by a sigmoidal function, and an analytically exact 
treatment to the transformation of waves is developed based on the time-dependent wave equation~(\ref{Maxwell3}).

\subsection{Smooth change of the dielectric permittivity}
\label{smooth.change}

\begin{figure}
\centering
\includegraphics[width=0.48\textwidth]{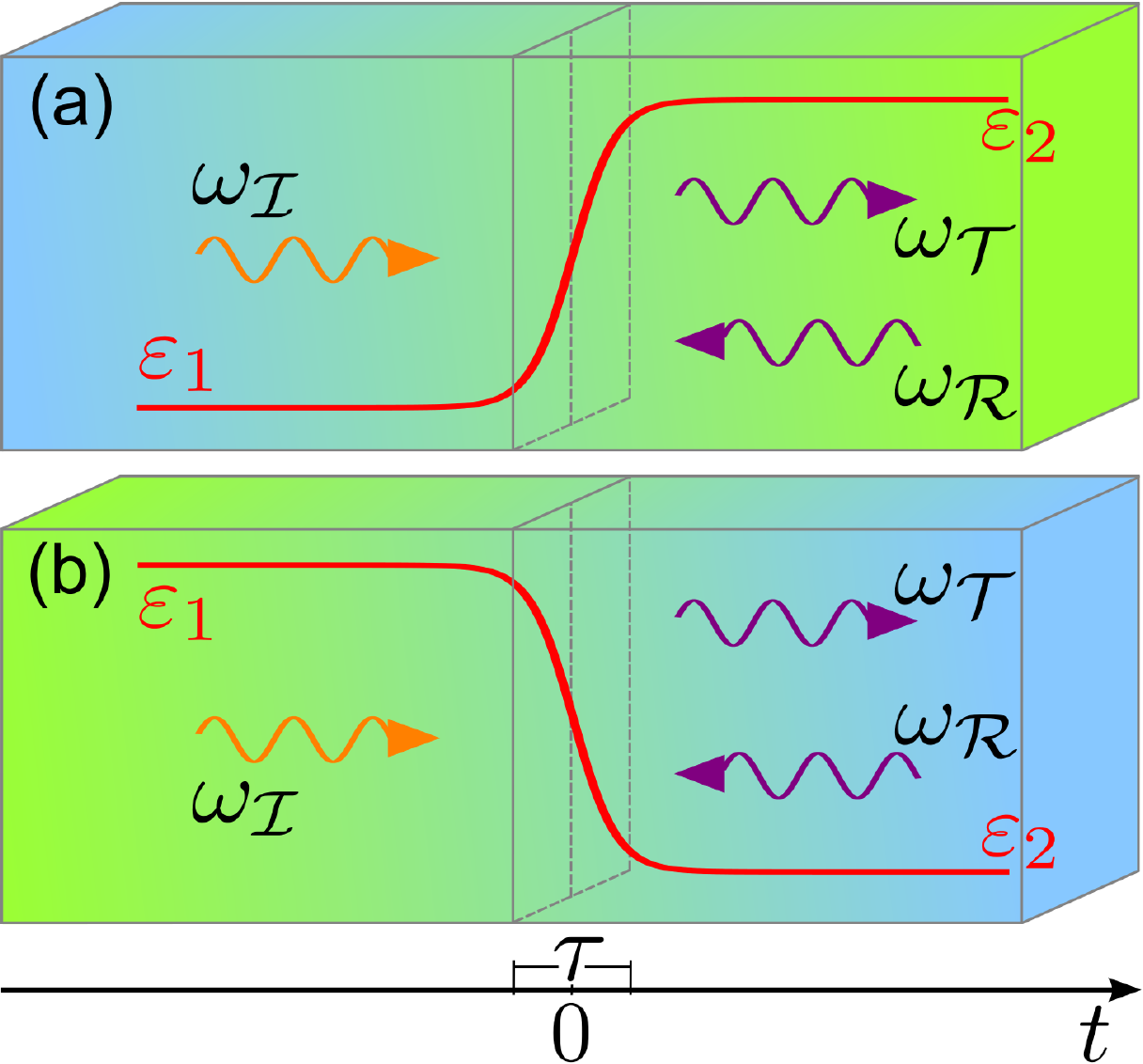}
\caption{Transformation of electromagnetic waves in spatially 
homogeneous yet time-dependent dielectric media when the 
permittivity either increases (a) or decreases (b) smoothly during some finite transition
period $\tau$. Smooth changes are modelled via Eq.~(\ref{permittivity}).} 
\label{figure3}
\end{figure}

In a more realistic case, the dielectric permittivity, instead of an abrupt variation, often experiences a change
during a finite transition period $\tau$. In order to account for such a finiteness, it 
is no longer sufficient to consider continuity 
conditions~(\ref{continuity.cond}): we need to solve the time-dependent 
wave equation~(\ref{Maxwell3}) where the sudden change of the permittivity is
replaced by a judiciously chosen and smoothly time-varying function, as depicted in Fig.~\ref{figure3}.
We may therefore model the switching of the permittivity by a 
phenomenological sigmoidal function
\begin{eqnarray}
\label{permittivity}
	 \varepsilon \left( t \right) & = &
	 \frac{\varepsilon_1 \varepsilon_2 \left( {1 + e^{t/\tau}} \right)}
	 {\varepsilon_2 +  \varepsilon_1 e^{t/\tau}} \, ,	
\end{eqnarray}
to assure that the asymptotic values 
$ \varepsilon_1$ (for $t < 0$) and $ \varepsilon_2$ (for $t > 0$) are recovered
when $\tau \rightarrow 0$.
Choosing this shape of the permittivity, we derive  
an exact second order linear differential equation 
from Eq.~(\ref{Maxwell3})
\begin{eqnarray}
\label{Maxwell4}
	\zeta \left( 1 - \zeta \right) \frac{d^2  {\cal D} }{d \zeta^2}
	+ \left( 1 - \zeta \right) \frac{d \, {\cal D} }{d \zeta}
	+ \left( \frac{\alpha + \beta}{\zeta} - \alpha \right) {\cal D} \,\, = \,\, 0 \, ,
\end{eqnarray}
where a new variable is introduced, $\zeta \equiv - e^{t/\tau}$, which
converges with $t \rightarrow - \infty$. The constant parameters
\begin{eqnarray}
\label{notations1}
	\alpha \, \equiv \, \frac{c^2 k^2 \tau^2}{\varepsilon_2} \, = \, \omega_{\cal T}^2 \tau^2 \, , \,\,
	\beta \, \equiv \, \frac{c^2 k^2 \tau^2}{\varepsilon_1} - \alpha  \, = \, 
	\left( \omega_{\cal I}^2 -  \omega_{\cal T}^2 \right) \tau^2  \, ,
\end{eqnarray}
expressed also by means of 
the initial $\omega_{\cal I}$ and transmitted (transformed) $\omega_{\cal T}$
frequencies,
carry information about the wave and the impact of the medium upon it.

Equation~(\ref{Maxwell4}) has a singularity at $\xi = 0$, which can be removed 
by making the replacement
\begin{eqnarray}
\label{Ansatz2}
	{\cal D} \left( \xi \right) & = & \xi^\nu \, {\cal F} \left( \xi \right) \, ,
\end{eqnarray}
where $\nu$ is a constant and, in general, complex number, while ${\cal F}$ represents an analytical function of $\xi$
and describes the time-dependent dynamics of the wave. 
The ansatz~(\ref{Ansatz2}) we advocate here amounts to reducing Eq.~(\ref{Maxwell4}) to the conventional form
\begin{eqnarray}
\label{Hypergeom.eq}
	\xi \left( 1 - \xi \right) \frac{d^2 {\cal F}}{d \xi^2} + \left( c - \left( a + b + 1 \right) \xi \right)
	\frac{d \, {\cal F}}{d \xi} - a b \, {\cal F} & = & 0 \, ,
\end{eqnarray}
where the constant parameters 
\begin{eqnarray}
\label{notations2}
	a \, \equiv \, \nu - i \sqrt{\alpha} \, , \,\,
	b \, \equiv \, \nu + i \sqrt{\alpha} \, , \,\,
	c \, \equiv \, 2 \nu + 1 
\end{eqnarray}
are introduced for the sake of brevity. Equation (\ref{Hypergeom.eq}) has an exact solution expressed in 
terms of the hypergeometric function, 
${\cal F} \left( \xi \right) = {\cal C} F \left( a, b, c, \xi \right)$ with ${\cal C}$
being a constant, provided that
$\nu^2 \equiv - \alpha - \beta = - \omega_{\cal I}^2 \tau^2$ to ensure the convergence of the solution
when $\xi = - e^{t/\tau} \rightarrow 0$~\cite{Gradshteyn:00}. This means that 
the solution is valid for negative values of $t$ since $\tau > 0$. 
Furthermore, we exploit Eqs.~(\ref{Ansatz1}) and (\ref{Ansatz2}) in order to construct the 
explicit form of the electric displacement in the interval $t<0$
\begin{eqnarray}
\nonumber
	D_{t < 0} \left( \bm r , t \right)
	\,\, = \,\,  e^{i \bm k \cdot \bm r} {\cal D}_{t < 0} \left( t \right) 
	\,\, = \,\, {\cal C} e^{i \bm k \cdot \bm r} \left(-1\right)^\nu e^{\nu t/\tau} F \left(a , b, c, -e^{t/\tau} \right) \, .
\end{eqnarray}
The asymptotic behaviour of this function defines the constant number ${\cal C}$ and the sign of $\nu$.
Given that $F \left( a , b , c , \xi \rightarrow 0 \right) \rightarrow 1$, the anticipated initial wave~(\ref{initial.wave})
can be gained if 
\begin{eqnarray}
\label{asympt.neg.t}
	{\cal C} \, = \, D_{{\cal I} 0} \left( - 1 \right)^{- \nu} \, , \quad
	\nu \, = \, - i \omega_{\cal I} \tau \, ,
\end{eqnarray}
such that the electric displacement itself takes the final form
\begin{eqnarray}
\label{solution.neg.t}
	D_{t < 0} \left( \bm r , t \right)
	& = &  D_{{\cal I} 0} \, e^{i \left( \bm k \cdot \bm r - \omega_{\cal I} t \right)} F \left(a , b, c, -e^{t/\tau} \right) \, .
\end{eqnarray}
This holds for all times $t<0$ for the given `rate of change' of the dielectric permittivity~(\ref{permittivity}).

The solution~(\ref{solution.neg.t}) does also contain the transformed waves~(\ref{refl.trans.waves})
that in turn incorporates the two -- reflected and transmitted -- waves. 
For $t > 0$, however, the hypergeometric function diverges because of
the argument $ - e^{t/\tau}$. We therefore need to employ its symmetry properties 
to circumvent this divergence. By building a new convergent variable  $ 1/\xi = - e^{ - t/\tau}$ at $t > 0$,
in fact, one can re-write the solution~(\ref{solution.neg.t}) as~\cite{Gradshteyn:00}
\begin{eqnarray}
\nonumber
	D_{t > 0} \left( \bm r , t \right) & = & D_{{\cal I} 0} \, e^{i \left( \bm k \cdot \bm r - \omega_{\cal I} t \right)}
\\[0.2cm]
\nonumber
	& \times &
	\left[ \frac{\Gamma \left( c \right) \Gamma \left( b - a \right)}{\Gamma \left( b \right) \Gamma \left( c - a \right)} 
	e^{-a t / \tau}
	F \left(a , a + 1 - c, a + 1 - b, -e^{ - t/\tau} \right)
	\right.
\\[0.2cm]
\label{solution.pos.t}
	& + &
	\left. \frac{\Gamma \left( c \right) \Gamma \left( a - b \right)}{\Gamma \left( a \right) \Gamma \left( c - b \right)} 
	e^{-b t / \tau}
	F \left(b , b + 1 - c, b + 1 - a, -e^{ - t/\tau} \right) 
	\right] \, , \quad\quad
\end{eqnarray}
that represents a superposition of two waves with time-varying complex `amplitudes'
expressed in terms of the hypergeometric functions.
In Eq.~(\ref{solution.pos.t}), the presence of the two exponentials indicates, generally, the frequency conversion 
owing to the terms $ - i \omega_{\cal I} t - a t / \tau = - i \omega_{\cal R} t$
and $ - i \omega_{\cal I} t - b t / \tau = - i \omega_{\cal T} t$~[cf. Eq.~(\ref{notations2})], which
mean that the non-stationary medium serves as a frequency transformer for waves~\footnote{
Similar effects related to frequency conversion, modulation and shift
attracted continuously growing interest since the past decades~\cite{Nerukh:12}.
See also Refs.~\cite{Hayrapetyan:16,Manzoni:15} for the frequency modulation in parity-time symmetric 
non-stationary structures and for the frequency conversion in the quantum regime, respectively.}. 
In order to disentangle these frequency- and time-dependent terms from the hypergeometric function (that is,
to ensure that they occur \textit{only} in the exponentials),
we shall consider the dynamics of the transformed waves long after the permittivity experiences the change.
In this limiting case, since $F \left( a , b , c , 1/ \xi \rightarrow  0 \right) \rightarrow 1$,  
Eq.~(\ref{solution.pos.t}) can be re-written as
\begin{eqnarray}
\nonumber
	D_{t > 0} \left( \bm r , t \right) & \cong & D_{{\cal I} 0} \, e^{i \bm k \cdot \bm r }
\\[0.2cm]
\label{solution.pos.t.inf}
	& \times &
	\left[ \frac{\Gamma \left( c \right) \Gamma \left( b - a \right)}{\Gamma \left( b \right) \Gamma \left( c - a \right)} 
	e^{i \omega_{\cal T} t }
	 + 
	 \frac{\Gamma \left( c \right) \Gamma \left( a - b \right)}{\Gamma \left( a \right) \Gamma \left( c - b \right)} 
	e^{- i \omega_{\cal T} t }
	\right] \, , \quad\quad
\end{eqnarray}
which shows explicitly the occurrence of two counter-propagating waves with modified amplitudes determined
by the $\Gamma$-functions. As we expect 
the solution~(\ref{solution.pos.t.inf}) to take the form~(\ref{refl.trans.waves}) for $t \rightarrow \infty$, 
further comparison of Eqs.~(\ref{refl.trans.waves}) and (\ref{solution.pos.t.inf}) gives 
\begin{eqnarray}
\label{amplitude.ratio}
	\frac{D_{{\cal R} 0}}{D_{{\cal I} 0}} \,\, = \,\,
	 \frac{\Gamma \left( c \right) \Gamma \left( b - a \right)}{\Gamma \left( b \right) \Gamma \left( c - a \right)} \, , \,\,
	\frac{D_{{\cal T} 0}}{D_{{\cal I} 0}} \,\, = \,\,
	\frac{\Gamma \left( c \right) \Gamma \left( a - b \right)}{\Gamma \left( a \right) \Gamma \left( c - b \right)} \, .
\end{eqnarray}
These expressions define the $\tau$-dependent amplitudes of the reflected and transmitted waves normalized to  
the amplitude of the initial wave (see also Eqs.~(\ref{notations1}) and (\ref{notations2})).
In the limiting case $\tau \rightarrow  0$, moreover, our general treatment 
confirms the results of Refs.~\cite{Avetisyan:78,Mkrtchyan:10,Morgenthaler:58,Mend:03}
as $\Gamma \left( z \right) \rightarrow 1 / z$ for $\left| z \right| \rightarrow 0$, but disagrees with 
Ref.~\cite{Mend:02} where the authors derive incorrect coefficients
despite using the same continuity conditions as Eq.~(\ref{continuity.cond})~\footnote{This mistake is 
later corrected in the subsequent paper~\cite{Mend:03}.}.

Solutions~(\ref{solution.neg.t})-(\ref{solution.pos.t.inf}) show how the smooth change of the dielectric permittivity~(\ref{permittivity})
manifests itself in the electric displacement, and therefore, affects the dynamical properties 
of electromagnetic waves in a non-stationary medium. Since these solutions -- established already for the entire time axis $t \gtrless 0$ --
depend explicitly on the switching duration (but not the mechanism!) of the refractive index, the transformation of waves and
their energy exchange with the non-stationary medium will also depend on the switching duration $\tau$
and the conversed frequency $\omega_{\cal T}$. To demonstrate this, we employ
the recurrence relation and Euler's reflection formula for the $\Gamma$-function
\begin{eqnarray}
\nonumber
	\Gamma \left( \vartheta + 1 \right) & = & \vartheta \, \Gamma \left( \vartheta \right) \, ,
\\[0.2cm]
\nonumber
	\left| \Gamma \left( i \vartheta \right) \right|^2 & = &
	\Gamma \left( i \vartheta \right) \Gamma \left( - i \vartheta \right) \,\, = \,\,
	\frac{\pi}{\vartheta \sinh \left( \pi \vartheta \right)} \, ,
\end{eqnarray}
use Eqs.~(\ref{reflection-transmission}) and (\ref{amplitude.ratio}),
we finally obtain the $\tau$-dependent reflectivity and transmittivity   
\begin{eqnarray}
\label{reflection.smooth}
	{\cal R} & = &  \frac{\omega_{\cal T}^2}{\omega_{\cal I}^2} 
	\frac{\sinh^2 \left( \pi \left( \omega_{\cal I} - \omega_{\cal T} \right) \tau \right)}
	{\sinh \left( 2 \pi \omega_{\cal I} \tau \right) \sinh \left( 2 \pi \omega_{\cal T} \tau \right)} \, , 
\\[0.2cm]
\label{transmission.smooth}
	{\cal T} & = &  \frac{\omega_{\cal T}^2}{\omega_{\cal I}^2} 
	\frac{\sinh^2 \left( \pi \left( \omega_{\cal I} + \omega_{\cal T} \right) \tau \right)}
	{\sinh \left( 2 \pi \omega_{\cal I} \tau \right) \sinh \left( 2 \pi \omega_{\cal T} \tau \right)} \, .
\end{eqnarray}
This is one of the main results of this work.
By generalizing the Fresnel-type formulae~(\ref{reflection.sudden}) and (\ref{transmission.sudden})
from sudden ($\tau \rightarrow 0$) to smooth ($\tau > 0$) transition of the permittivity, the 
expressions~(\ref{reflection.smooth})-(\ref{transmission.smooth}) describe the 
energy transport of an electromagnetic wave when propagating through 
a spatially homogeneous yet smoothly time-varying medium
for the specific time dependence (\ref{permittivity}). 
We will utilize formulae (\ref{reflection.smooth})-(\ref{transmission.smooth}) 
in the next section in order to discuss the energy balance
between waves and non-stationary media, which depends on the switching duration and
reveals the amplification and attenuation of waves.

\section{Results and discussion}
\label{res.disc}

\begin{table}
  \caption{Comparison of transformation of electromagnetic and sound waves waves in smoothly time-dependent media
  described by either the refractive index ($n_1$, $n_2$) or the mass density ($\rho_1$, $\rho_2$) and the sound 
  velocity distributions ($V_1$, $V_2$). ${\cal T}^{\scriptstyle (s)}$ and ${\cal R}^{\scriptstyle (s)}$ correspondingly 
  represent the energy flux transmission and 
  reflection coefficients for sound waves, i.e., the sound reflectivity and transmittivity.}
    \label{table2}
  \begin{tabularx}{\textwidth}{lcc}
    \hline\hline 
     & Electromagnetic waves                                  & Sound waves~\cite{Hayrapetyan:13} \\
    \hline 
    General, $\tau$-dependent    &{\large \cmark}  Eqs.~(\ref{reflection.smooth})-(\ref{transmission.smooth})
                                                   & {\large \cmark} Eqs.~(53)-(54) of \cite{Hayrapetyan:13} \\
    Fresnel-type formulae            &                    &\\[0.2cm]
    Energy balance           & {\large \cmark} $n_1 \neq n_2$ 
                                        & {\large \cmark} $\rho_1 \neq \rho_2$, $V_1 \neq V_2$ \\
                                        & [Eq.~(\ref{energy-exchange.smooth})]    & [Eq.~(56) of \cite{Hayrapetyan:13}] \\[0.2cm]
    Amplification of waves & {\large \cmark} $n_1 > n_2$
                                        & {\large \cmark} $\rho_1 \gtrless \rho_2$, $V_1 \gtrless V_2$ \\[0.2cm]
    Attenuation of waves   & {\large \cmark} $n_1 < n_2$
                                        & {\large \xmark} \\[0.2cm]
    Energy difference        & {\large \cmark} ${\cal T} - {\cal R} = n_1^2 / n_2^2 $ 
    				      & {\large \cmark} ${\cal T}^{\scriptstyle (s)} - {\cal R}^{\scriptstyle (s)} = 1$ \\
                                        & [Eq.~(\ref{difference})]   & [Eq.~(57) of \cite{Hayrapetyan:13}] \\
    \hline\hline
  \end{tabularx}
\end{table}

For both the sudden and smooth changes of the permittivity, the wave exchanges
energy with the medium [cf. Eq.~(\ref{energy-exchange.sudden})] and, as a result, is either 
amplified or attenuated. For a smooth change, that occurs during the finite period~$\tau$, the
sum of the reflectivity~(\ref{reflection.smooth}) and the transmittivity~(\ref{transmission.smooth})
takes the form
\begin{eqnarray}
\label{energy-exchange.smooth}
	{\cal R} + {\cal T} \,\, = \,\,
	\frac{\omega_{\cal T}^2}{ \omega_{\cal I}^2}
	\left( \frac{\tanh \left( \pi \omega_{\cal I} \tau \right)}
	{2 \tanh \left( \pi \omega_{\cal T} \tau \right)} +
	\frac{\tanh \left( \pi \omega_{\cal T} \tau \right)}
	{2 \tanh \left( \pi \omega_{\cal I} \tau \right)} \right) \,\, \neq \,\, 1 \, .
\end{eqnarray}
As the expression in the parentheses is always larger
than unity, we immediately see that the wave is either amplified 
or attenuated depending on whether the conversed frequency is increased or
decreased (as compared to the initial one), or else, the 
refractive index is decreased or increased [cf.~Eq.~(\ref{conversion})].  
A similar situation also holds for the sudden change of the medium,
as described by Eq.~(\ref{energy-exchange.sudden}). Being one of the
main results of this paper, the expression (\ref{energy-exchange.smooth})
generalizes Eq.~(\ref{energy-exchange.sudden}) to explicitly include
the switching duration $\tau$ and shows the universal nature of the 
amplification and attenuation of electromagnetic waves also in 
the case of smooth change of the medium. This is in contrast
with sound waves which due to their inherent structure are only amplified, irrespective, whether
the mass density and sound velocity distributions
increase or decrease as functions of time.
In Table~\ref{table2}, a comparison is made in terms of generalized Fresnel-type formulae
for the electromagnetic and sound waves.

Another interesting feature can be obtained from Eqs.~(\ref{reflection.smooth})-(\ref{transmission.smooth})
if we calculate the difference of the transmittivity and reflectivity 
\begin{eqnarray}
\label{difference}
	{\cal T} - {\cal R} \,\, = \,\, \frac{\omega_{\cal T}^2}{\omega_{\cal I}^2}
	\,\, = \,\, \frac{n_1^2}{n_2^2} \, ,
\end{eqnarray}
which is \textit{in}dependent of $\tau$ and maintains the same form 
as that for the sudden change~\cite{Mkrtchyan:10}. 
While the wave travels from optically rarer to denser medium ($n_2 > n_1$),
the transmitted wave carries an energy flux smaller than the sum of
the energy fluxes in the reflected and initial waves, and vice versa
for $n_2 < n_1$: the energy flux of the transmitted wave surpasses
that of the two other waves. This is again 
in contrast to sound waves where the transmitted wave carries an energy flux
exactly equal to the sum of the fluxes of the reflected and initial waves~\cite{Mkrtchyan:10,Hayrapetyan:13}
[cf.~Table~\ref{table2}]. 

\begin{figure}
\centering
\includegraphics[width=0.98\textwidth]{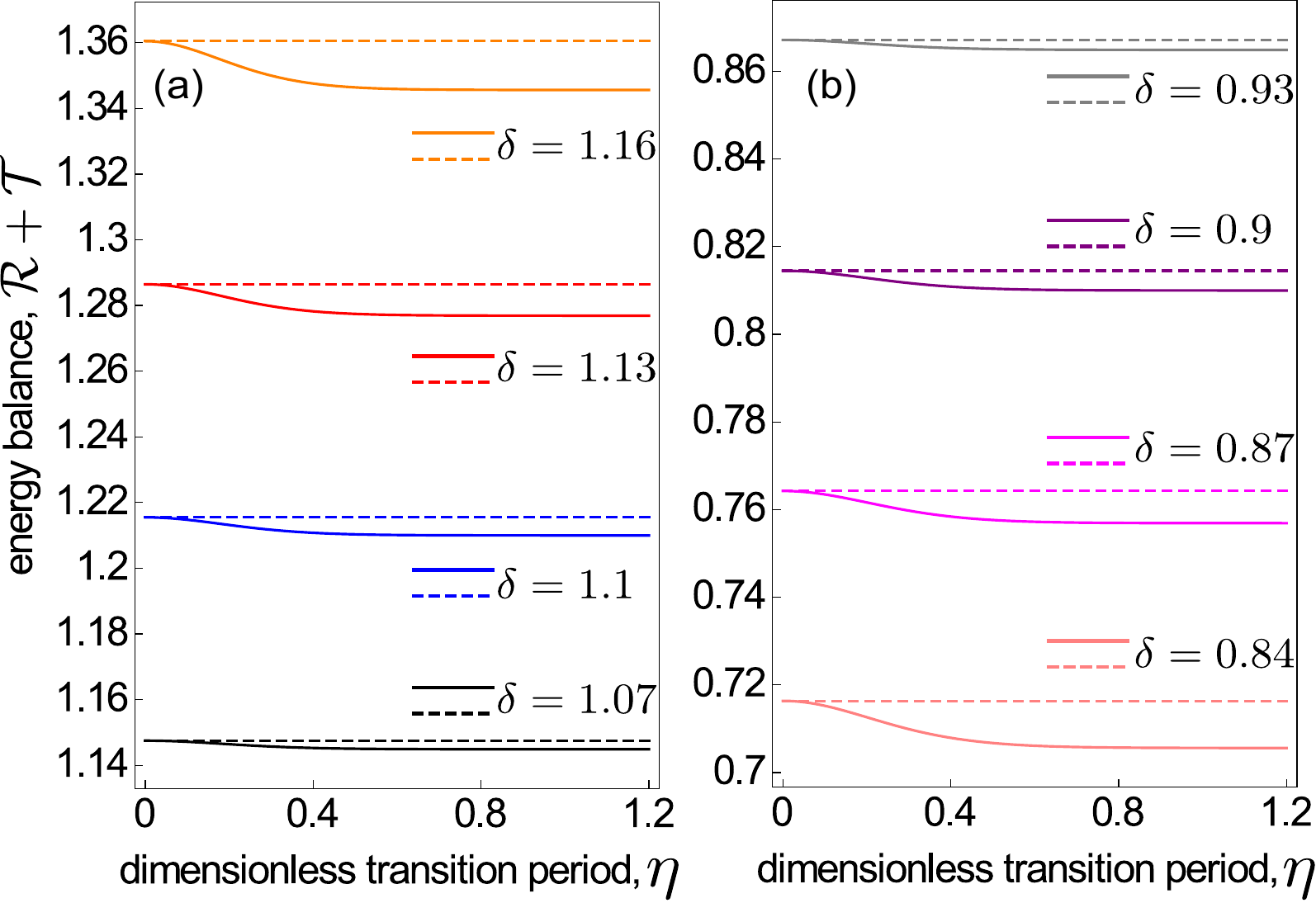}
\caption{Amplification (a) and attenuation (b) of electromagnetic waves in 
suddenly (dashed lines) and smoothly (solid curves) changing
dielectric media. Comparison is made between different  
ratios $\delta$ of refractive indices before and after the change;
$\delta > 1$ and $\delta < 1$ correspond to the decrease (a) and increase (b)
of the refractive index. The case $\delta = 1$, when no energy exchange occurs,
is not shown on the figure.} 
\label{figure4}
\end{figure}
To better perceive the energy balance between the electromagnetic wave
and the non-stationary medium, let us re-write the expressions~(\ref{energy-exchange.smooth}) and (\ref{energy-exchange.sudden})  
in the dimensionless form
\begin{eqnarray}
\label{exchange.smooth-for.plots}
	{\cal R} + {\cal T} & = &
	\delta^2
	\left( \frac{\tanh \left( \pi \eta \right)}
	{2 \tanh \left( \pi \delta \eta \right)} +
	\frac{\tanh \left( \pi \delta \eta \right)}
	{2 \tanh \left( \pi \eta \right)} \right) \,\, \neq \,\, 1 \, ,
\\[0.2cm]
\label{exchange.sudden-for.plots}
	\left( {\cal R} + {\cal T} \right) |_{\eta \rightarrow 0} & = &
	\frac{\delta}{2} \left( 1 + \delta^2 \right) \,\, \neq \,\, 1 \, .
\end{eqnarray}
Here, $\eta = \omega_{\cal I} \tau$ represents the dimensionless transition period, while 
the parameter
$\delta \equiv \omega_{\cal T} / \omega_{\cal I} = n_1 / n_2$ shows the
ratio between the refractive indices before and after the change, so that 
$\delta $ equals to unity when the tuning of the refractive index is 
``switched off''. 
Figure~\ref{figure4} demonstrates the energy balance between 
the wave and the medium as a function of $\eta$ 
for different values of $\delta$ as well as for both the sudden (dashed lines) and smooth (solid curves)
changes of the dielectric permittivity. The fact that the sum of the reflectivity 
and transmittivity is either greater (Fig.~\ref{figure4}(a)) or less
(Fig.~\ref{figure4}(b)) than the unity is a signature of the wave amplification and attenuation, respectively. 
This change in the sum of the energy fluxes of transformed waves is quantified by means of the ratio $\delta$
between the refractive indices.
As seen, the larger the ratio $\delta$ or $1/\delta$, the stronger is the 
amplification or attenuation of the wave. The inclination of curves for the smooth change
is more pronounced as $\delta$ increases (decreases) from unity due to 
the variation $d \left( {\cal R} + {\cal T} \right) / d \eta \approx - \pi^2 \delta \left( \delta^2 - 1 \right)^2 \eta / 3$
for small $\eta$. Particularly, the increase of $7$ \% in $\delta$ 
corresponding to the decrease of the refractive index ($\delta = 1.07$)
gives rise to the change of $\sim 15$ \% in the amplification (${\cal R} + {\cal T} \approx 1.15$) [cf. 
black curves in Fig.~\ref{figure4}a]. Whereas the decrease of $7$~\% in $\delta$ results in 
the decrease of $\sim 13$ \% in the sum of the reflectivity and transmittivity and leads to an attenuation 
of waves (${\cal R} + {\cal T} \approx 0.87$)
[cf. grey curves in Fig.~\ref{figure4}b]. 
\begin{figure}
\centering
\includegraphics[width=0.48\textwidth]{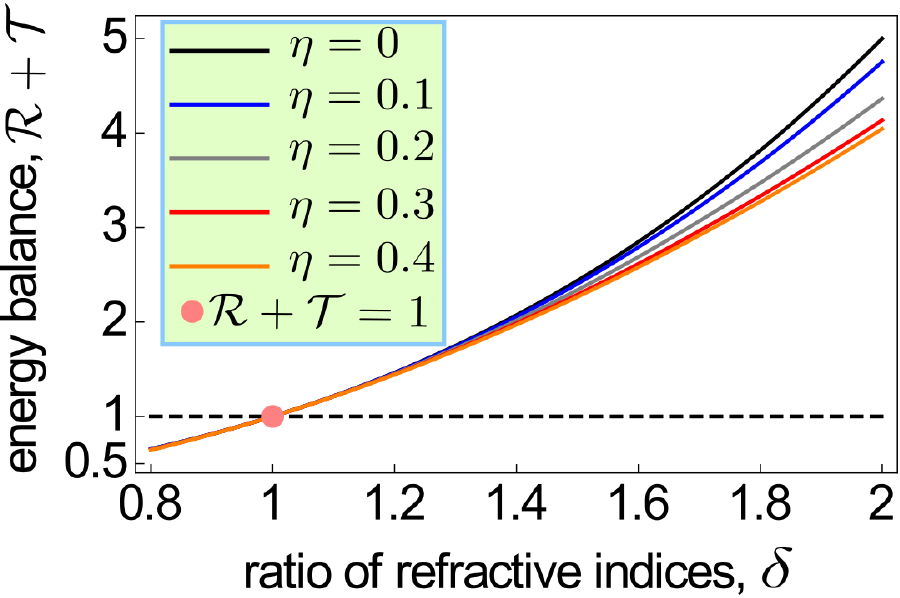}
\caption{Energy balance versus ratio between refractive indices
before and after the change of the medium for different values of the
dimensionless time $\eta$. The pink point illustrates 
the area where the energy of the wave is conserved, $\delta = 1$.} 
\label{figure5}
\end{figure}
Such an asymmetry between 
the increase and decrease of energy fluxes is due to the fact that Eqs.~(\ref{exchange.smooth-for.plots}) 
and~(\ref{exchange.sudden-for.plots}) do not remain symmetric under the interchange 
$\delta \leftrightarrow 1/\delta$ (or $n_1 \leftrightarrow n_2$).
As expected, moreover, when the transition period is much less than the period of the initial wave
($\tau \ll 2 \pi / \omega_{\cal I}$)
both curves for sudden and smooth changes
at given $\delta$ merge to each other as $\tanh \left( \pi \delta \eta \right)/\tanh \left( \pi \eta \right) \rightarrow \delta$
for $\eta \rightarrow 0$. Figure~\ref{figure5}, in turn, illustrates the dependence
of the energy balance on the ratio~$\delta$ between the refractive indices
for selected values of the dimensionless time~$\eta$. As the dimensionless time
approaches its asymptotic value ($\eta \rightarrow 0$), 
representing the sudden change of the permittivity,
the difference between curves decreases. The 
dependence on $\eta$ 
is markedly pronounced in the domain of amplification, which is again
due to the $\delta \leftrightarrow 1/\delta$ asymmetry of the
energy balance~(\ref{exchange.smooth-for.plots}).

Thus, apart from being a frequency transformer, 
the non-stationary medium acts as a `source or sink of energy' for electromagnetic waves. 
The prediction of an increase and decrease of normalized energy flux 
for the wave, which can correspondingly be 
interpreted as an amplification and attenuation due to an energy exchange 
with a medium, can be tested experimentally, if it is possible to disentangle the
energy transport through the dielectric from its internal energy.
A microscopic theory should be developed in order to explicitly reveal 
the source of energy, as in our phenomenological approach 
the time-dependent permittivity characterizes only a `net' structure, which can be generated, for instance,
by laser pulses in photonic systems~\cite{Mondia:05,Yuece:12}.

\section{Conclusion}
\label{conclusion}

We have derived an analytically exact theory to describe the propagation
and transformation of electromagnetic waves in spatially homogeneous 
yet smoothly time-varying dielectric structures.
The emphasis has been put on exploring how the finite transition period
$\tau$ for the dielectric permittivity influences the dynamical properties
of waves, such as the energy (flux) exchange between waves and non-stationary media 
and the conversion of frequencies of transformed waves. 
The exchange is shown to lead to the $\tau$-dependent amplification or attenuation of waves
correspondingly linked to the wave propagation from optically denser 
to rarer medium or vice versa. 
We have provided a detailed comparison between predictions of our
generalized theory and those of the sudden change approximation.
The peculiar differences in transformations of electromagnetic
and sound waves in smoothly-varying media are also pointed out.
Being manifestations of the temporal 
inhomogeneity, both the energy exchange and the
transformation of frequencies can be tested experimentally should
the switched dielectric permittivity follow the sigmoidal shape,
as shown in Fig.~\ref{figure3}. However,  
a rigorous study of ubiquitous processes of relaxation following the switching of the
refractive index would constitute modification of the sigmoidal change, and therefore
correction to the reflectivity~(\ref{reflection.smooth}) and 
transmittivity~(\ref{transmission.smooth}).

Although our results are valid for a wide range of 
frequencies (from radio frequency up to ultraviolet) and for various types of dielectric media,
a generalized approach is needed to simultaneously account for smoothly time-dependent
dielectric permittivity and magnetic permeability, especially relevant for 
studying transformation of waves in non-stationary 
plasmas~\cite{Kalluri:92,Kalluri:93,Lee:98,Kalluri:09} and magnetoelectric systems~\cite{Zhang:15}.
Such a general study would enable one to resolve the
debate about various ways of deriving the reflection and transmission
coefficients in suddenly changing media~\cite{Xiao:14,Bakunov:14}.
In recent years, moreover, controlling waves in both space and time domains has raised considerable 
interest~\cite{Mosk:12,Thyrrestrup:14}. Simultaneous
investigation of space- and time-dependent transformation of waves 
will lead to an intriguing `interplay' of energy and momentum
exchange between waves and spatially inhomogeneous and non-stationary media.

\section*{Acknowledgements}

AGH acknowledges useful discussions with J\"org Evers at early stage of this work
and thanks Willem Vos and Georgios Ctistis for their insightful 
comments on the experimental realization of time-varying refractive index.

\end{document}